\documentstyle[12pt]{article}
\begin{document}
\begin{flushright}
Preprint:~IMSc/970727 \\
hep-ph/9709212
\end{flushright}

\vskip 1cm

\begin{center}
{\large \bf Perspectives in High Energy Physics}

\vskip .7cm

{\it G. Rajasekaran} \\
{\it Institute of Mathematical Sciences} \\
{\it Madras - 600 113} \\
{\it India.}
\end{center}

\vskip 2cm

\centerline{\bf Abstract}

\vskip .5cm

\baselineskip=14pt

A broad survey of High Energy Physics (HEP) both
within as well as beyond the Standard Model is
presented emphasizing the unsolved problems. Inspite of the 
spectacular success of
the Standard Model, there is a serious crisis facing the field. The
importance of
research on new methods of acceleration that can resolve
this crisis by taking us to superhigh energies is
stressed. We briefly review the status of HEP in India and offer
suggestions for the future.

\vskip 4cm

\begin{center}
Invited talk presented at XII HEP Symposium, Guwahati,
India \\
26 December '96 ~-~ 1 January 97.
\end{center}

\newpage

\noindent {\bf 1. History}

\vskip .5cm

The major events which culminated in the construction
of the Standard Model of High Energy Physics are
presented in Table 1 in chronological order. Using
nonabelian gauge theory with Higgs mechanism, the
electroweak (EW) theory was already constructed in
1967, although it attracted the attention of most
theorists only after another four years, when it was
shown to be renormalizable. The discovery of
asymptotic freedom of non abelian gauge theory and the
birth of QCD in 1973 were the final inputs that led
to the full standard model.

\vskip .5cm

On the experimental side, the discovery of scaling in
deep inelastic scattering (DIS) which led to the
asymptotic free QCD and the discovery of the neutral
current which helped to confirm the electroweak theory
can be regarded as crucial experiments. To this list,
one may add the polarized electron-deuteron experiment
which showed that $SU(2) \times U(1)$ is the correct gauge
group for electroweak theory, the discovery of
gluonic jets in electron-positron annihilation
confirming QCD and the discovery of $W$ and $Z$ in
1983 that established the electroweak theory. The
experimental discoveries of charm, $\tau$, beauty and 
top were fundamental for the concrete 3-generation
standard model.

\vskip.5cm

However, note the blank after 1973 on the theoretical
side. Theoretical physicists have been working even
after 1973 and experiments also are being done. But
the tragic fact is that none of the bright ideas
proposed by theorists in the past 23 years has
received any experimental support. On  the other side,
 none of the experiments done since 1975 has made an
independent discovery. They have only been confirming
the theoretical structure completed in 1973. It is
clear that if such a situation persists for long, it
may become difficult to continue to be optimistic
about the future of high energy physics. We shall take up this point in
Sec.3.

\newpage

\begin{center}
{\bf Table 1. ~History of the Standard Model}

\vskip .5cm

\begin{tabular}{|lll|lll|} \hline
\multicolumn{3}{|c|}{\bf Theory} & \multicolumn{3}{|c|}{\bf Experiment}
\\ \hline
& & & & & \\
& 1954 & Nonabelian & & & \\
& & ~~~~gauge fields & & & \\
& & & & & \\
1960 & & & & & 1960 \\
& 1964 & Higgs mechanism & & & \\
& 1967 & EW Theory & 1968 & Scaling in DIS & \\
& & & & & \\
1970 & & & & & 1970 \\
& 1971 & Renormalizability & & & \\
& & ~~~ of EW Theory & & & \\
& 1973 & Asymptotic freedom & 1973 & Neutral current & \\
& & \multicolumn{1}{c|}{$\rightarrow$ QCD} & & & \\
& & & 1974 & Charm & \\
& & & 1975 & $\tau$ - lepton & \\
& & & 1977 & Beauty& \\
& & & 1978 & $\overrightarrow{e} d$ expt & \\
& & & 1979 & gluonic jets & \\
& & & & & \\
1980 & & & & & 1980 \\
& & & 1983 & W,Z & \\
& & & & & \\
1990 & & & & & 1990 \\
& & & 1994 & top & \\
& & & & & \\ \hline
\end{tabular}
\end{center}

\vskip .5cm

\noindent {\bf 2. Perspectives and Highlights of the Symposium}

\vskip .5cm

The standard model based on the gauge group $SU(3)
\times SU(2) \times U(1)$ describes {\it all} of
presently known High Energy Physics. How well the standard model fits the
data, was
reviewed in the talks of Gautam Bhattacharyya, Somnath Ganguli and Atul Gurtu.
This is the peak where we have reached. From here we can survey the view
either below us (i.e. within the standard model) or
above us (i.e. beyond the standard model). Possible
topics in either view are the following :-

\vskip .5cm

\baselineskip=14pt
\noindent {\it\underline{Within the standard model}}

\vskip .5cm

QCD and hadronic physics 

Higgs and symmetry breaking

Neutrinos 

Generation problem

CP, axion etc

\vskip .5cm

\noindent {\it\underline{Beyond the standard model}}

\vskip .5cm

Preons

Grand Unification

Supersymmetry and Supergravity

Higher Dimensional Unification 

Superstrings

\vskip .5cm

\indent \indent Let me first dispose of the view below
the standard model.

\vskip .5cm

\noindent {\it\underline{QCD and hadronic physics}}

\vskip .5cm

\indent \indent Here the questions are the following :

\begin{itemize}
\item[(i)] Can we establish QCD to be the correct
theory of strong interaction?
\item[(ii)] Can colour confinement be proved?
\item[(iii)] Can hadron spectrum be calculated?
\item[(iv)] Can hadron scattering be calculated?
\item[(v)] Do glue balls exist?
\item[(vi)] Does quark-gluon plasma exist?
\end{itemize}

\baselineskip=14pt

Ten years ago I talked on ``Perspectives in HEP'' (Ref. : Proceedings of
VIII High
Energy Physics Symposium, Calcutta, 1986, p.399). The above list of topics and
questions is in fact taken from that talk. Have the questions raised at
that time,
been answered? In the following, I shall enclose the quotations from the
1986 talk
as ``~~~~~~~~~~~~~~~~~~~~~~~~".

\vskip .5cm

``Unfortunately at the present moment, the answer to all
these questions is negative. Answer to the first
question will depend on the answers to the next three
questions. Lattice-gauge-theorists are working hard on
these problems. Here a word of
caution may be appropriate, concerning the numerical
calculation of hadronic properties such as their
masses and couplings. It must be remembered that these
properties of hadrons have been calculated earlier
more than once in the history of high energy physics -
first within the analytic $S$ matrix and bootstrap
approach and later in quark potential models. Each
time success was claimed. The real test of any
numerical calculation in hadronic physics must be the
prediction of {\it a new number or a new phenomenon in
the area of strong interaction}, which is then
confronted with experiment. Until that is achieved,
success cannot be claimed. After all, what is the
sense of using expensive computer time to calculate
the masses of the hadrons, when these can be obtained
with much greater accuracy, by looking up the
excellent Particle Data Tables?'' Although the main point of these critical
statements still stands, one has to admit that important new developments have
occurred. {\it Asit De} gave a very lucid review of these and claimed
that lattice
QCD results are just starting to enter Particle Data Tables. This is good news!

\vskip .5cm

``In the absence of a clean check of QCD in the realm of
the dirty hadrons, the existence of glue balls or the
transition of hadronic matter into quark-gluon plasma
would be a direct and striking confirmation of QCD.
But distinguishing glue balls from flavour-singlet
quark balls has not proved a clean job. Let us hope
that the imminent heavy-ion collisions will produce
the eagerly awaited quark-gluon plasma and that the
plasma will announce its arrival with a clean signal''. Heavy-ion
collisions have
occurred, but people are still searching for clean signals of QGP! {\it
C.P.Singh}
reviewed the current status of this field. 

\vskip.5cm

What about continuum QCD? Light-front QCD appears to be a promising
approach and
progress in it was reported by {\it Harindranath}. A scholarly review on
thermal
field theory was given by {\it Samir Mallik}, who pointed out that the
infra-red
problem for finite temperature QED has been solved by {\it Indumathi}.
The status of
perturbative QCD and the structure function of the proton as revealed by
HERA was
reviewed by {\it Dilip Choudhury} and {\it Rahul Basu}.

\vskip .5cm

\noindent {\it Higgs and symmetry breaking}

\vskip .5cm

``Is Higgs the correct mechanism of electroweak symmetry
breaking? There are claims from the axiomatic side
that $\lambda \phi^4$ theory may be an inconsistent
theory. Should Higgs mechanism be replaced by some
other nonperturbative dynamical symmetry breaking?
Inspite of much effort, we have not progressed much
towards an understanding of dynamical symmetry
breaking. Experiments being planned in the Tev region
may reveal either the presence of Higgs bosons or a
new type of strong interactions in the electroweak
sector. In either case, we will have an exciting time''. {\it
S.R.Choudhury} showed how
the triviality of $\lambda \phi^4$ theory combined with consistency can
be used to
yield bounds on Higgs mass and {\it D.P.Roy} described the ongoing
searches for the Higgs
boson.

\vskip .5cm

\noindent {\it Neutrinos, Generations, CP, Axion etc}

\vskip .5cm

``Are the neutrinos massless? If not, what are their
masses and mixing angles? The recent elegant
explanation of the solar neutrino puzzle by {\it
resonant} neutrino oscillations (the
Mikheyev-Smirnov-Wolfenstein effect) must be noted.
This explanation needs confirmation by independent
experiments such as that proposed by Raghavan and
Pakvasa (1987). Here one perhaps has a powerful tool
for pinning down neutrino masses and mixing
angles''. The atmospheric neutrino puzzle has now joined the solar
neutrino puzzle
and both indicate neutrino oscillations. Neutrino physics has grown into an
important field. Data from the new generation of neutrino detectors
(Super-Kamioka,
SNO and Borexino) are eagerly awaited. Also, long-base-line terrestrial
neutrino
experiments are being planed.

\vskip .5cm

``How may generations of quarks and leptons exist and what
fixes this number?

Of the various options within the standard model for
explaining CP violation, which is the correct one?

Is Peccei-Quinn symmetry and axion the correct cure for
the catastrophe of strong CP violation in QCD? If so,
where is the axion?''

\vskip .5cm

``On all these questions, enormous amount of theoretical work
has been done, but no memorable results have come out.
So, most theorists have gone out of the standard model
to make a living. This is not surprising, for this is
what the theorists have been always doing. We did not
solve all the problems of atomic physics before moving
on to nuclear physics, nor did we understand nuclear
physics fully before inventing a new field called
particle physics and moving into it. After reaching a
peak we do not set up our permanent quarters there;
we climb to the next peak. So, we move on to \ldots
beyond the standard model.'' I then went on to describe Preons, SUSY \& SUGRA,
Higher Dimensions and finally Strings, which contained the following remark. 

\vskip .5cm

``Further, search for consistent theories of even more complicated objects than
strings, for instance, membranes, lumps \ldots etc must continue. Any
reported ``No
go'' theorem in this context need not be regarded as a permanent barrier.
Remember,
without the invention of SUSY and acceptance of higher dimensions, even string
theories would suffer a ``No go'' theorem. There will be discovered other things
which will make the theories of membranes, lumps and even objects extending to
higher dimensions consistent.''

\vskip .5cm

This is what has happened now. We are witnessing a Second Revolution in String
Theory which has converted String Theory itself into a Theory of Branes (objects
extending to arbitrary number of  dimensions). 

\vskip .5cm

Following are a few highlights that dealt with ``Beyond the
Standard Model'', in this symposium.

\vskip .5cm

\noindent {\it Supersymmetry}

\vskip .5cm

Probir Roy, D.P.Roy and Ananthanarayanan presented comprehensive reviews of
supersymmetric theories. We still await their experimental discovery.

\vskip .5cm

\noindent {\it String theory}

\vskip .5cm

{\it Sunil Mukhi} gave a stimulating talk on the recent developments. 
Using the web of {\it duality} they are catching a rich harvest of
interconnections between various string theories and they are already getting a
glimpse of a so-called $M$-theory which may be the fundamental source of
all string
theories, membrane theories etc. 

\vskip .5cm

If string theory is the correct theory of Quantum Gravity it should help us to
understand black holes better and the recent developments have achieved
this. It is
the understanding of the solitons and D-branes of string theory that has
contributed
to this development and {\it Dabholkar} dealt with this topic.

\vskip .5cm

After listening to any talk on this Second Revolution in string theory, I
feel so
envious of my younger colleagues who are making such a fantastic progress
in this
difficult and highly competitive subject. (I wish I were 20 years younger!)

\vskip .5cm

\noindent {\it Two applications of string theory}

\vskip .5cm

\noindent {\it (a) Proton stability}

\vskip .5cm

The problem of catastrophically fast proton decay ($\tau_p \sim 10^{-5} sec)$ in
supersymmetric theories, which is due to the existence of colour triplet
scalars in
these theories, is not yet solved. Conservation of $R$-parity is a
possible solution
and a few other solutions are technically possible, but not compelling.
No deeper
theoretical reason for proton stability has been found. {\it Jogesh Pati}
argued
that the real solution may require superstrings. Hopefully, this would
provide the
deeper reason. 

\vskip .5cm

\noindent {\it (b) CP violation}

\vskip .5cm

In an interesting talk, {\it David Bailin} sought CP violation in the orbifold
compactification of 10-dimensional heterotic strings. It may be possible to
incorporate CP as a geometrical transformation in a higher-dimensional
theory and 
hence its violation may have a geometrical origin.

\vskip .5cm

\noindent {\it Dualized standard model}

\vskip .5cm

In a beautiful work, {\it Tanmay Vachaspati} has shown how the standard model
could be dualized. He starts with $SU(5)$ and breaks it down to a version
of $SU(3)
\times SU(2)  \times U(1)$. The most remarkable aspect of his work is that no
fermions are put by hand. The solitonic monopoles that arise in the theory have
precisely the same magnetic charge as the electric charges on the quarks
and leptons
of the standard model. So, if we make the proper identification, the quarks and
leptons can be generated as solitons! This is certainly {\it a bolt from
the blue} and
deserves further study.

\vskip .5cm

\noindent {\it Topological quantum field theory}

\vskip .5cm

{\it Romesh Kaul} described how the QFT framework (which we use to
describe HEP) can be
used to reveal the topological properties of 3 and 4 manifolds.  Thus QFT has
enough power to move the frontiers of Modern Mathematics too! In
particular, duality
in cohomological field theory leads to an almost trivial calculation of
the famous
Donaldson invariants in 4-D, which are in turn related to instantons.
Since 4 is the
number of physical dimensions of space-time in which we live and since Donaldson
invariants are related to the infinite number of differential structures
that have
been proved to exist only in 4 dimensions, all this mathematics may have
profound
consequences for physics!

\vskip .5cm

\noindent {\bf 3. Does HEP have a future?}

\vskip .5cm

We now return to the blanks in discovery mentioned in Sec.1. The blanks have
remained inspite of the tremendous activity in HEP in the past two decades.
The biggest loophole in standard model is the omission of gravitation, the
most important force of nature. Hence, it is now recognized that {\it Quantum
Gravity (QG) is the next frontier of HEP}, and that {\it the true
fundamental scale of physics is the Planck energy 10$^{19}$ Gev}, which is
the scale of QG.

\vskip .5cm

We are now probing the TeV (10$^3$ GeV) region.
One can see the vastness of the domain one has to cover before
QG is incorporated into physics. In their attempts to probe this domain
of $10^3 - 10^{19} GeV$, theoretical physicists have invented many ideas
such as supersymmetry, supergravity, hidden dimensions etc and based on
these ideas, they have constructed  many beautiful theories, the
best among them being the superstring theory (or,
$M$-theory, its recent incarnation), which may turn out
to be the correct theory of QG. 

\vskip .5cm

But, Physics is not theory alone. Even beautiful theories have to be
confronted with experiments and either confirmed or thrown out. Here we
encounter a serious crisis facing HEP. In the next 10-15 years,
new accelerator facilities with higher energies such as the Large Hadron
Collider ($\sim 10^4$ GeV) or the Linear Electron Collider will be built
and so the prospects for HEP in the immediate future appear to be
bright. Beyond that period, the accelerator route seems to be closed
because known acceleration methods cannot take us beyond about $10^5$
GeV. 

\vskip .5cm

It is here that one turns to hints of new physics from
Cosmology, Astrophysics \& Nonaccelerator Experiments. 
Very important hints about neutrinos, dark matter etc. have come from
Astrophysics \& Cosmology. Nonaccelerator experiments on proton decay, 
neutrino masses, double beta decay and 5$^{th}$ force are important since they
provide us with indirect windows on superhigh energy scales.

\vskip .5cm

In spite of the importance of astroparticle physics \& nonaccelerator
experiments, these must be regarded as only our first and preliminary
attack on the unknown frontier. {\it These are only hints!} Physicists
cannot remain satisfied with hints and indirect attacks on the superhigh
energy frontier. {\it So, what do we do?}

\vskip .5cm

As already mentioned, the outlook is bleak, because known acceleration
methods cannot take us far.

\vskip .5cm

To sum up the situation :- \ There are many interesting fundamental
theories taking us to the Planck scale and even beyond, but unless the
experimental barrier is crossed, these will remain only as Metaphysical
Theories.

\vskip .5cm

It follows that either, {\it new ideas of acceleration have to be
discovered} or, {\it there will be an end to HEP by about 2010 A.D.}

\vskip .5cm

It is obvious what route physicists must follow.  We have to discover new
ideas on
acceleration. By an optimistic extrapolation of the growth of accelerator
technology
in the past 60 years, one can show that even the Planckian energy of
$10^{19}$ GeV
can be reached in the year 2086. (See my Calcutta talk). But, this is
possible only
if newer methods and newer technologies are continuously invented.

\vskip .5cm

Some of the ideas being pursued are laser beat-wave method, plasma wake
field accelerator, laser-driven grating linac, inverse free electron
laser, inverse Cerenkov acceleration etc. What we need are a hundred
crazy ideas. May be, one of them will work. Lawrence's discovery of the
cyclotron principle is not the end of the road. 

\vskip .5cm

\noindent {\bf 4. Status of HEP in India and Suggestions for the future}

\vskip .5cm

\noindent {\it Theory}

\vskip .5cm

There is extensive activity in HEP theory in the
country, spread over TIFR, PRL, IMSc, SINP, IOP, MRI,
IISc, Delhi University, Panjab University, BHU, NEHU,
Guwahati University, Hyderabad Univesity, Cochin
University, Viswabharati, Calcutta University,
Jadavpur University, Rajasthan University and a few
other Centres. Research is done in almost all the
areas in the field, as any survey will indicate.

\vskip .5cm

Theoretical HEP continues to attract the best students
and as a consequence its future in the country appears
bright. However, it must be mentioned that this
important national resource is being underutilized.
Well-trained HEP theorists are ideally suited to teach
any of the basic components of physics such as Quantum
Mechanics, Relativity, Quantum Field Theory,
Gravitation and Cosmology, Many Body Theory or
Statistical Mechanics and of course Mathematics, since all these ingredients go
to make up the present-day HEP Theory. Right now, most
of these bright young theoretical physicists are
seeking placement in the Research Institutions. Ways
must be found so that a larger fraction of them can be
absorbed in the Universities. Even if just one of them
joins each of the 200 Universities in the country,
there will be a qualitative improvement in physics
teaching throughtout the country. This will not happen unless the young
theoreticians gain a broad perspective in the topics mentioned above and train
themselves for teaching-cum-research careers. Simultaneously, the
electronic communication facilities linking the
Universities among themselves and with the Research
Institutions must improve. This will solve the
frustrating isolation problem which all the University
Departments face.

\vskip .5cm

\noindent {\it Experiment}

\vskip .5cm

Many Indian groups from National Laboratories as well
as Universities (TIFR, VECC, IOP, Delhi, Panjab, Jammu
and Rajasthan Universities) have been participating in
3 major international collaboration experiments :-

\vskip .5cm

\begin{itemize}
\item $L3$ experiment on $e^+~e^-$ collisions at LEP (CERN)
\item $D\phi$ experiment on $\bar{p}p$ collisions at the
Tevatron (Fermilab)
\item WA93 \& 98 experiments on heavy-ion collisions at CERN.
\end{itemize}

\vskip .5cm

Highlights of the Indian contribution in these experiments
were presented in this symposium.

\vskip .5cm

As a result of the above experience, the Indian groups
are well poised to take advantage of the next
generation of colliders such as LEP2 and the LHC.
Already the Indian groups have joined the
international collaboration in charge of the CMS which
will be one of the two detectors at LHC. It is also
appropriate to mention here that Indian engineers and
physicists will be contributing towards the
construction of LHC itself.

\vskip .5cm

Thus, the only experimental program that is pursued in
the country is the participation of Indian groups in
international accelerator based experiments. This is
inevitable at the present stage, because of the nature
of present-day HEP experiments that involve
accelerators, detectors, experimental groups and
financial resources that are all gigantic in
magnitude.

\vskip .5cm

While our participation in international
collaborations must continue with full vigour, at the
same time, for a balanced growth of experimental HEP,
we must have in-house activities also. Construction of
an accelerator in India, in a suitable energy range
which may be initially 10-20 GeV and its utilization
for research as well as student-training well provide
this missing link.

\vskip .5cm

In view of the importance of underground laboratories
in $\nu$ physics, monopole search, $p$ decay etc, the
closure of the deep mines at KGF is a serious loss.
This must be at least partially made up by the
identification of some suitable mine and we must
develop it as an underground laboratory  for 
nonaccelerator particle physics.

\vskip .5cm

Finally, it is becoming increasingly clear that known
methods of acceleration cannot take us beyond tens of
TeV. Hence in order to ensure the continuing vigour of
HEP in the 21st century, it is absolutely essential to
discover new principles of acceleration. Here lies an
opportunity that our country should not miss! I have been respeatedly
emphasizing for the past ten years that we must form a small group of
young people
whose mission
shall be to discover new methods of acceleration.

\vskip .5cm

To sum up, a 4-way program for the future of
experimental HEP in this country is suggested :

\vskip .5cm

\begin{enumerate}
\item A vigorous participation of Indian groups in
international experiments, accelerator-based as well
as non-accelerator-based.
\item Construction of an accelerator in this country.
\item Identification and development of a suitable
underground laboratory for nonaccelerator particle physics.
\item A programme for the search of new methods of
acceleration that can take HEP beyond the TeV
energies.
\end{enumerate}

\vskip .5cm

\noindent {\bf Acknowledgement and Apology}

\vskip .5cm

I thank Dilip Choudhury for the invitation to give this talk and excellent
hospitality at Guwahati. I apologize to those whose contributions could not be
highlighted in my talk.
\end{document}